\documentclass[aps,prb,twocolumn,showpacs,preprintnumbers,amsmath,amssymb,superscriptaddress]{revtex4}%

\usepackage{graphicx}%
\usepackage{dcolumn}
\usepackage{amsmath}
\usepackage{color}
\usepackage{bm}

\begin{document}

\title{Synthesis, crystal structure and chemical stability of the  superconductor FeSe$_{1-x}$}

\author{E.~Pomjakushina}
\email[]{ekaterina.pomjakushina@psi.ch}
\affiliation{Laboratory for Developments and Methods, PSI, 5232
Villigen, Switzerland}
\author{K.~Conder}
\affiliation{Laboratory for Developments and Methods, PSI, 5232
Villigen, Switzerland}
\author{V.~Pomjakushin}
\affiliation{Laboratory for Neutron Scattering, ETHZ \& PSI, 5232
Villigen, Switzerland}
\author{M.~Bendele}
\affiliation{Physik-Institut der Universit\"{a}t Z\"{u}rich,
Winterthurerstrasse 190, 8057 Z\"urich, Switzerland}
\affiliation{Laboratory for Muon Spin Spectroscopy, PSI, 5232
Villigen, Switzerland}
\author{R.~Khasanov}
\affiliation{Laboratory for Muon Spin Spectroscopy, PSI, 5232
Villigen, Switzerland}

\def\figsize{10cm}

\date{\today}

\begin{abstract}

We report on a comparative study of the crystal structure and the
magnetic properties of FeSe$_{1-x}$ ($x = 0.0 - 0.15$)
superconducting samples by neutron powder diffraction and
magnetization measurements. The samples were synthesized by two
different methods: a ''low-temperature`` one using powders as a
starting material at $T\simeq700^{\rm o}$C  and a
''high-temperature`` method
  using  solid pieces of Fe and Se at $T\simeq1075^{\rm o}$C.
The effect of a starting (nominal) stoichiometry on the phase purity
of the obtained samples,  the superconducting transition temperature
$T_{\rm c}$, as well as the chemical stability of FeSe$_{1-x}$ at
ambient conditions were investigated. It was found that in the Fe-Se
system a stable phase exhibiting superconductivity at $T_{\rm c}
\simeq 8K$ exists in a narrow range of selenium concentration
(FeSe$_{0.974\pm 0.005}$).
\end{abstract}


\maketitle

\section{Introduction}

The discovery of Fe-based superconductors has attracted considerable
attention to the pnictides. Superconductivity is detected  now in
various pnictide families as {\it e.g.} the single-layer
ReO$_{1-x}$F$_x$FeAs (Re=La, Ce, Pr, Nd, Sm, Gd, Tb, Dy, Ho and Y),
\cite{Kamihara08,Wang08,Ren08,Sefat08,Chen08,Wen08,Yang08} the
double-layer MFe$_2$As$_2$ (M=Ba, Sr, and Ca),
\cite{Rotter08,Rotter08a,Chen08a,Ni08,Torikachvili08} the oxygen
free single-layer LiFeAs \cite{Wu08,Pitcher08,Tapp08} {\it etc.} The
common structural feature of all these materials is the Fe-As layers
consisting of a Fe square planar sheet tetrahedrally coordinated by
As. Recently, superconductivity with a transition temperature of
$T_{\rm c}\simeq8$~K was discovered in $\beta-$FeSe$_{1-x}$ with
PbO-structure. \cite{Hsu08} This compound also has a Fe square
lattice with Fe atoms tetrahedrally coordinated by Se  similar to
the structure of FeAs layers in the single- and the double-layer
pnictides. In this respect FeSe$_{1-x}$, consisting of the
''superconducting`` Fe-Se layers only, can be considered as a
prototype of the known families of Fe-As based superconductors and,
consequently, is a good model system to study mechanisms leading to
the occurrence of superconductivity in this new class of materials.

As  is stated in the literature, there are two different routs to
synthesize superconducting FeSe$_{1-x}$.
The first one uses Se and Fe powders as the starting material and is
performed in sealed silica tubes  at 400-700$^{\rm o}$C.
\cite{Hsu08} Hereafter we call it  the ''low-temperature`` synthesis
(LTS). This method, however, was shown to result in samples with
relatively high amount of impurities. According to
Ref.~\onlinecite{Hsu08}, FeSe$_{1-x}$ with $x=0.18$ was found to
consist  of a superconducting phase and a small amount of elemental
selenium, iron oxide and iron silicide (reaction product with silica
ampoule). For a higher average selenium content ($x=0.12$),  some
amount of  hexagonal (NiAs-type) FeSe phase was detected in addition
to impurity phases listed above. The superconducting transition
temperature was found to be at $\simeq$ 8~K, being independent of
the initial Se content.
The second procedure proposed in the  recent work of McQueen et al.
\cite{McQueenPRB2009}starts from Fe pieces and Se shot. The Fe and
Se pieces sealed in silica ampule, were first hold at 750$^{\rm o}$C
(3-5 days), then heated up to 1075$^{\rm o}$C (3 days) followed by a
fast decrease down to 420$^{\rm o}$C and quenched. The synthesis was
completed by additional annealing of the sample (sealed in a new
ampoule) at 300-500$^{\rm o}$C  followed by quenching.
 Superconductivity was found  to exist only in a very narrow range of
stoichiometry. For FeSe$_{0.99}$ (Fe$_{1.01}$Se) magnetization
measurements showed $T_{\rm c}\sim 8.5K$ whereas $T_{\rm c}$ for
FeSe$_{0.98}$ (Fe$_{1.02}$Se)  decreased down to $\sim$ 5K and went
to zero (at least down to 0.6K) for FeSe$_{0.97}$
(Fe$_{1.03}$Se).\cite{McQueenPRB2009} Hereafter we call this
procedure  the ''high-temperature`` synthesis (HTS). In comparison
with LTS, the samples prepared by HTS
 do not contain iron oxide impurities. \cite{McQueenPRB2009}

Surprisingly, the FeSe$_{1-x}$ samples synthesized by LTS and HTS
techniques were found to be rather different. Indeed, in LTS samples
 superconductivity was found  in a rather extended range of Se
content (at least up to
$x=0.18$),\cite{Hsu08,MargadonnaChemComm2008} while for the HTS
 superconductivity was detected only in a very narrow region
corresponding to $0.01\leq x\leq0.025$.\cite{McQueenPRB2009} In
addition, McQueen {\it et al.}\cite{McQueenPRB2009} reported, that
below 300$^{\rm o}$C the tetragonal FeSe$_{1-x}$ converts into a
hexagonal (NiAs-type) phase which is not superconducting. Therefore,
quenching from temperatures above 300$^{\rm o}$C was used for
synthesizing HTS samples. On the other hand, no special care for
fast cooling of LTS samples needs to be
taken.\cite{Hsu08,MargadonnaChemComm2008}  In order to resolve these
controversies  we performed comparative studies of the
superconducting FeSe$_{1-x}$ samples synthesized by both methods
(LTS and HTS) described above. We have improved a method of
synthesis using powder starting materials and investigated the
effect of  stoichiometry on the phase purity of the obtained samples
and its superconducting transition temperature. Based on the neutron
powder diffraction data we have  revised  Fe-Se concentration phase
diagram proposed by Okamoto. \cite{OkamotoJPhEq1991}

\section{Experimental details}

\begin{figure}[htb]
\begin{center}
\leavevmode
\includegraphics[width=1\linewidth]{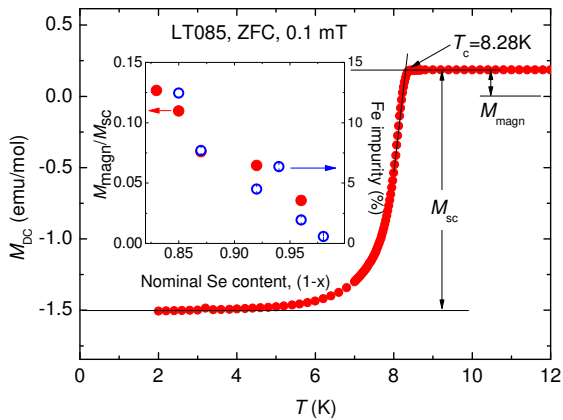}
\end{center}
 \caption{(Color online) Temperature dependencies of
the DC magnetization ($M_{DC}$, zero-field cooling, 0.1~mT) of LT085
sample. The superconduting transition temperature $T_{\rm c}$, the
superconducting ($M_{\rm SC}$) and the magnetic ($M_{\rm magn}$)
responses of the sample are determined as shown in the figure. The
inset shows  dependencies of $M_{\rm magn}/M_{\rm SC}$ and   Fe
impurity concentration as a function of the nominal Se content. }
 \label{fig:Magn-LT085}
\end{figure}

AC and DC magnetization ($M_{AC}/M_{DC}$) measurements were
performed using Quantum Design PPMS and MPMS magnetometers at
temperatures ranging from 2 to 300~K. The AC field amplitude and the
frequency were 0.1~mT and 1000~Hz, respectively. The DC
magnetization experiments were performed after zero-field cooling
and field cooling  the samples at $\mu_0H=0.1$~mT. The
superconducting transition temperature $T_{\rm c}$ was determined as
an intersection of the linearly extrapolated
$M_{AC}(T)$[$M_{DC}(T)$] with the $M=const$ line (see
Fig.~\ref{fig:Magn-LT085}).

Neutron powder diffraction (NPD) experiments were carried out at
the SINQ spallation source at the Paul Scherrer Institute (PSI,
Switzerland) using the high-resolution diffractometer for thermal
neutrons HRPT\cite{hrpt} (the neutron wavelengths
$\lambda=1.494$~{\AA} and 1.155~{\AA}).  The refinements of the
crystal structure parameters were done using {\tt FULLPROF}
program,\cite{Fullprof} with the use of its internal tables for
neutron scattering lengths.

\section{Results and discussion}

\subsection{Sample synthesis}

Two types of samples  using the LTS  and the HTS  procedure were
prepared. Samples of a nominal composition FeSe$_{0.85}$,
FeSe$_{0.87}$, FeSe$_{0.92}$, $\rm FeSe_{0.96}$, FeSe$_{0.98}$, and
FeSe (LT085, LT087, LT092, LT096, LT098 and LT100) were prepared by
a solid state reaction similar to that described in
Refs.~\onlinecite{Hsu08} and \onlinecite{MargadonnaChemComm2008}.
The cold pressed mixtures of Fe and Se powders were sealed in quartz
ampoules and then heated up to 700$^{\rm o}$C followed by annealing
at 400$^{\rm o}$C. Powders of Fe and Se of a minimum purity of
99.99\% were used as starting materials.

Sample FeSe$_{0.94}$ (HT094) was synthesized similar to the route of
McQueen {\it et al.} \cite{McQueenPRB2009} - by solid state reaction
using pieces of Fe and Se of a minimum purity of 99.99\%. The sample
was heated in the sealed quartz ampoule up to 1075$^{\rm o}$C
followed by annealing at 400$^{\rm o}$C. In contrast to McQueen {\it
et al.}\cite{McQueenPRB2009} no quenching from high temperatures was
made. The sample was cooled down to the room temperature at a rate
of 200$^{\rm o}$C/h.

Note that for the samples synthesized by both LTS and HTS techniques
all the grindings/pelletizings were performed under  helium
atmosphere. The samples studied in the present work are listed in
the Table~\ref{Table:results}.

\subsection{Se content and impurity phases}
\begin{figure}[t]
\begin{center}
\leavevmode
\includegraphics[width=1\linewidth]{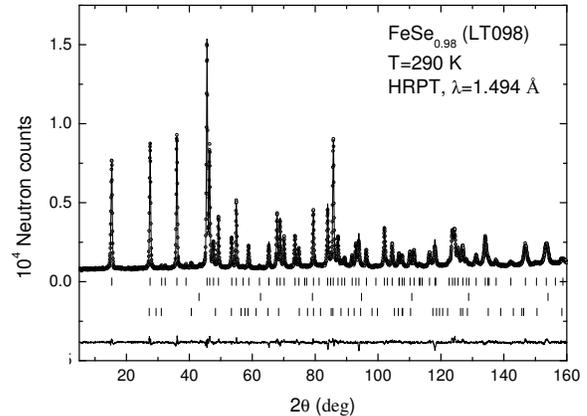}
\end{center}
 \caption{The Rietveld refinement pattern and difference plot of
the neutron diffraction data for the sample FeSe$_{0.98}$ (LT098) at
T=290~K measured at HRPT with the wavelength $\lambda=1.494$~{\AA}.
The rows of ticks show the Bragg peak positions for the main phase
and two impurity phases. The main tetragonal phase corresponds to
0.975(3) Se content. See text for details. }
 \label{fig:NPD}
\end{figure}

The chemical composition of the main (superconducting)  phase and
the  concentration of the impurity phases  were determined by means
of neutron powder diffraction. First we present the results obtained
for the FeSe$_{1-x}$ samples with nominal Se contents $0.85\leq 1-x
\leq 0.98$ (LT085--LT098 and HT094). Room temperature NPD
experiments show that all these samples contain the same  tetragonal
phase FeSe$_{1-x}$ (space group \emph{P4/nmm}) as a main phase. The
refined selenium occupancy (selenium stoichiometry) is about
$0.974(2)$ and  is {\it independent}  of the starting composition
and the route of the synthesis (LTS, HTS). A typical example of
Rietveld refinement of NPD data is shown in Fig.~\ref{fig:NPD} for
the LT098 sample. Impurity phases are hexagonal FeSe (space group
$P6_{3}mmc$) in a quantity of $\sim1$\% (molar \%) and Fe (space
group \emph{Im3m}). The amount of the metallic Fe was found to
decrease monotonically with increasing Se content from
$\simeq$12.5\% for FeSe$_{0.85}$ to $\simeq0.5$\% for FeSe$_{0.98}$.
Note, that for all the samples studied  the presence of any oxides
was not detected. The "cleanest" sample is LT098 which contains, in
total, less than 2\% of the secondary phases and has a nominal
composition FeSe$_{0.98}$, the same as is refined for the main
tetragonal phase FeSe$_{0.975(3)}$. The amount of impurity phases
found in the samples, the refined stoichiometry of the main
tetragonal phase and its unit cell parameters are listed in
Table~\ref{Table:results}.

\begin{table*}[t]
\caption[~]{\label{Table:results} Summary of the neutron powder
difraction (NPD) and magnetization results  for FeSe$_{1-x}$
samples prepared by LTS and HTS methods. }
\begin{center}

\begin{tabular}{|l|l|c|l|l|l|l|l|}
\hline sample & nominal  &  $T_{\rm c}$(K)& phase
content (molar \%) & unit cell parameters ({\AA}) \\
& composition & & & of the tetragonal phase \\
\hline \hline LT085a & $\rm FeSe_{0.85}$& - &$\rm FeSe_{0.994(11)}$ (\emph{P4/nmm}) 71.75$\pm$1.75\%& a=3.77413(14) \\
&&& Fe (\emph{Im3m}) 26.23$\pm$0.85\% & c=5.52141(31)\\
&&& $\rm FeSe$ ($P6_{3}/mmc$) 2.02$\pm$0.38\% & \\
\hline LT085 & $\rm FeSe_{0.85}$ & 8.28& $\rm FeSe_{0.963(5)}$ (P4/nmm) 86.38$\pm$0.98\%& a=3.77320(4)\\
&&& Fe (\emph{Im3m}) 12.46$\pm$0.33\% & c=5.52496(9)\\
&&& $\rm FeSe$ ($P6_{3}/mmc$) 1.16$\pm$0.18\%& \\
\hline LT087 & $\rm FeSe_{0.87}$& 8.34&  $\rm FeSe_{0.979(4)}$ (P4/nmm) 91.53$\pm$0.91\%& a=3.77280(4)\\
&&& Fe (\emph{Im3m}) 7.70$\pm$0.23\% & c=5.52303(8)\\
&&& $\rm FeSe$ ($P6_{3}/mmc$) 0.77$\pm$0.14\% & \\
\hline LT092 & $\rm FeSe_{0.92}$& 8.44&  $\rm FeSe_{0.976(4)}$ (P4/nmm) 94.50$\pm$0.89\%& a=3.77335(4)\\
&&& Fe (\emph{Im3m}) 4.50$\pm$0.21\% & c=5.52368(8)\\
&&& $\rm FeSe$ ($P6_{3}/mmc$) 1.00$\pm$0.13\% & \\
\hline HT094 & $\rm FeSe_{0.94}$& 8.21 &  $\rm FeSe_{0.977(3)}$ (P4/nmm) 92.91$\pm$0.70\%& a=3.77294(4)\\
&&& Fe (\emph{Im3m}) 6.36$\pm$0.16\% & c=5.52421(7)\\
&&& $\rm FeSe$ ($P6_{3}/mmc$) 0.73$\pm$0.09\% & \\
\hline LT096 & $\rm FeSe_{0.96}$& 8.43&  $\rm FeSe_{0.978(4)}$ (P4/nmm) 96.02$\pm$1.07\%& a=3.77338(5)\\
&&& Fe (\emph{Im3m}) 1.94$\pm$0.23\% & c=5.52415(11)\\
&&& $\rm FeSe$ ($P6_{3}/mmc$) 2.04$\pm$0.19\% & \\
\hline LT098 & $\rm FeSe_{0.98}$& 8.21&  $\rm FeSe_{0.975(3)}$ (P4/nmm) 98.31$\pm$0.59\%& a=3.77381(2)\\
&&& Fe (\emph{Im3m}) 0.57$\pm$0.05\% & c=5.52330(5)\\
&&& $\rm FeSe$ ($P6_{3}/mmc$) 1.12$\pm$0.08\% & \\
\hline LT100 & $\rm FeSe_{1}$& $\sim 8.0$&  $\rm FeSe_{0.968(3)}$ (P4/nmm) 83.03$\pm$0.61\%& a=3.77353(4)\\
&&& Fe (\emph{Im3m}) 0.46$\pm$0.05\% & c=5.52382(7)\\
&&& $\rm Fe_{7}Se_{8}$ ($P3_{1}21$) 16.51$\pm$0.28\% & \\
\hline \hline
\end{tabular}
\end{center}
\end{table*}

The results of the structural analysis were further confirmed by
magnetic susceptibility measurements. As follows from
Table~\ref{Table:results}, all FeS$_{1-x}$ samples ($0< x \leq
0.15$) have almost the same transition temperatures ($T_{\rm
c}\sim8.2-8.4$~K) and, consequently, very similar doping
(concentration of charge carriers). In addition, the paramagnetic
offset ($M_{magn}$), seen at $T>T_{\rm c}$ was found to decrease
with increasing Se content just following the dependence of Fe
impurity phase as the function of the nominal Se content $1-x$ (see
the inset in Fig.~\ref{fig:Magn-LT085}). Note that in
Ref.~\onlinecite{Khasanov08FeSe} the observation of the paramagnetic
offset at $T>T_{\rm c}$ as well as the static magnetic contribution
seen in zero-field muon-spin rotation experiments were attributed to
the presence of Fe impurities.

By increasing the Fe:Se ratio up to 1:1 the situation was
drastically changed. The Rietveld refinement of NPD data on FeSe
(LT100) sample reveals that the main tetragonal phase content is
substantially decreased down to $\simeq83.03$\%. The content and the
composition of the impurity phases were also changed: only 0.46\% of
Fe and, instead of a hexagonal NiAs-type  phase, $\simeq 16.51$\% of
the trigonal Fe$_7$Se$_8$ (space group $P3_{1}21$) was detected.
Magnetization experiments  also show the substantial decrease in the
superconducting fraction which was found to be of about 10\% at
$T=3$~K.

\begin{figure}[t]
\begin{center}
\leavevmode
\includegraphics[width=0.9\linewidth]{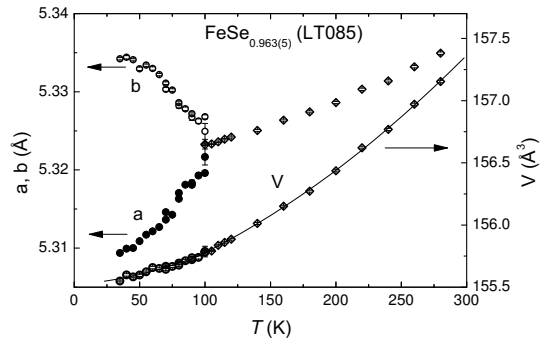}
\end{center}
 \caption{
$a,b$ unit cell parameters and unit cell volume $V$ as a function of
temperature. In the tetragonal phase the lattice constant is multiplied by $\sqrt{2}$.}
 \label{fig:abV}
\end{figure}

\begin{figure}[t]
\begin{center}
\leavevmode
\includegraphics[width=0.9\linewidth]{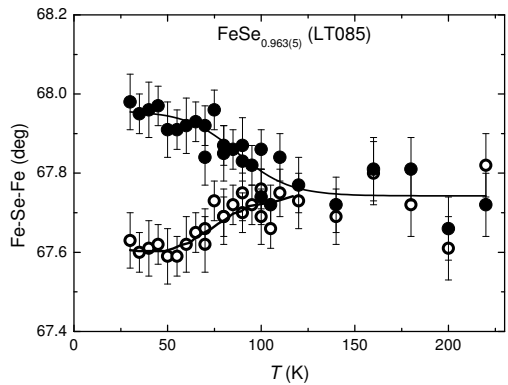}
\end{center}
 \caption{Fe-Se-Fe  bond angles as a function of temperature in $\rm FeSe_{0.963(5)}$.
 The refinements of the diffraction data were made assuming low symmetry phase (space group \emph{Cmma}).
 The high temperature crystal structure is tetragonal and both Fe-Se-Fe angles would be the same by symmetry
in $P4/nmm$ group.}
 \label{fig:angles}
\end{figure}

Studies of the crystal structure of the main phase as a function of
temperature were performed on the sample LT085 in the temperature
range 20-300 K on both cooling and heating. Figure~3 shows the
lattice constants $a$,$b$ and the unit cell volume as a function of
temperature. At
 temperature $T$=100K there is a transition from the tetragonal to
orthorhombic structure on cooling similar as observed in
Ref.~\onlinecite{MargadonnaChemComm2008}. The low temperature
structure is well refined in the space group $Cmma$. The building
block of the crystal structure is $\rm SeTe_4$ square pyramide with
Se-atom in the apex. In the high temperature phase the pyramide is
regular, whereas in the orthorhombic phase the neighboring Fe-Se-Fe
bond angles become different as shown in Fig.~4. The Se-Fe bond
length amounts to 2.386(2)~\AA\ at low temperature and it is not
changed at the transition. Neither temperature hysteresis nor the
unit cell volume jump were observed indicating that the transition
is of the second order type. The transition temperature in our
sample $\rm FeSe_{0.963(5)}$ (100~K) is different from the one
reported for  $\rm FeSe_{0.91}$ (70~K) \cite{MargadonnaChemComm2008}
that might be related to the different Se stoichiometry. However, as
described above our synthesis techniques always produce the main
tetragonal phase with approximately the same concentration with the
average value of about x=0.974.

\subsection{Phase diagram}

\begin{figure}[t]
\begin{center}
\leavevmode
\includegraphics[width=1.5\linewidth]{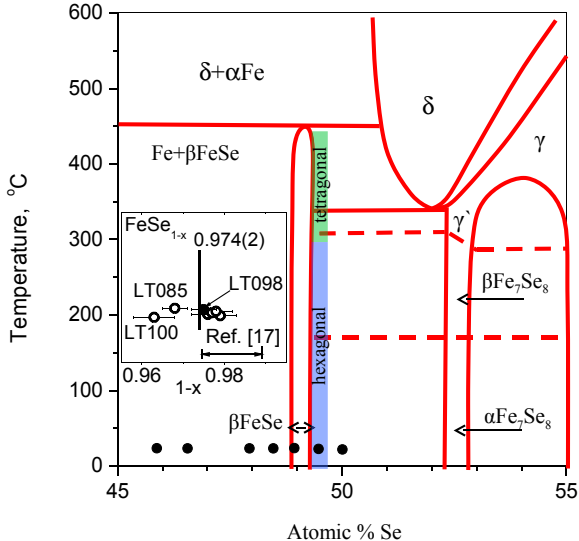}
\end{center}
\vspace{-2cm}
 \caption{(Color online) The Fe-Se phase diagram after Okamoto
Ref.~\onlinecite{OkamotoJPhEq1991} (lines) and McQueen {\it et al.}
Ref.~\onlinecite{McQueenPRB2009} (vertical colored stripe). The
black circles correspond to the nominal composition of the samples
studied in the present work. In the insert the refined selenium
stoichiometry (1-x) of the main superconducting phase found in all
the investigated samples are plotted together with the range of the
existence of non-stoichiometric FeSe$_{1-x}$ as proposed in
Ref.~\onlinecite{McQueenPRB2009}. See text for details. }
 \label{fig:Phase_diagram}
\end{figure}

In this section  stoichiometry of the main phase and the phase
composition of the studied samples are discussed based on  the
existing Fe-Se binary phase diagrams elaborated by Okamoto
\cite{OkamotoJPhEq1991} and McQueen {\it et al.}
\cite{McQueenPRB2009} Figure \ref{fig:Phase_diagram} shows a
combined phase diagram based on the previously published
data.\cite{McQueenPRB2009,OkamotoJPhEq1991} Lines correspond to the
part of the binary phase diagram reported in
Ref~\onlinecite{OkamotoJPhEq1991}. The stripe, centered at around
49.5 atomic percent of Se, corresponds to a range of existence of
tetragonal above $300^{\rm o}$C  and hexagonal FeSe$_{1-x}$
(NiAs-type) below $300^{\rm o}$C as proposed in
Ref.~\onlinecite{McQueenPRB2009}. The  circles correspond to the
samples with  different nominal Se content studied in this work. The
refined selenium stoichiometry (1-x) of the main superconducting
phase for all the investigated samples are plotted in the inset. The
existence range of the non-stoichiometric FeSe$_{1-x}$ as proposed
in Ref.~\onlinecite{McQueenPRB2009} is also shown. The average
stoichiometry of the superconducting phase was determined to be $\rm
FeSe_{0.974(2)}$ (the errorbar represents the statistical error).
The average Se-concentration is represented by the vertical line in
the insert and it is very close to that for the most pure sample
LT098 [$\rm FeSe_{0.975(3)}$], which is shown by the solid point in
the inset. It is worth to mention, that the samples showing the
largest deviation from the average stoichiometry (LT85 and LT100)
contain also a relatively large amount of the impurity phases.
Therefore the refined Se-occupancy may have some systematic error
for these samples. Most NPD measurements were performed with the
wavelength $\lambda=1.494$~\AA\ because it provides the optimal
conditions for refining the structure parameters  of the main pase
(large $q$-range) and determination of the impurity phases (good
resolution). To further check for the possible systematic error in
the Se-occupancy due to its correlation with the  atomic
displacement parameters  measurements of the most pure sample with
yet shorter wavelength $\lambda=1.155$~\AA were performed. The
refined Se-occupancy was found to be 0.980(3), implying that the
systematic error is smaller than 0.005.

From the data presented in Fig.~\ref{fig:Phase_diagram} the
following important statements emerge:
(i) the stability field of the superconducting \emph{tetragonal}
$\beta$-FeSe phase as proposed in Ref.~\onlinecite{McQueenPRB2009}
does not overlap with that reported  in
Ref.~\onlinecite{OkamotoJPhEq1991}. According to
Ref.~\onlinecite{McQueenPRB2009} the tetragonal phase exist only at high temperatures but
it is transformed to the hexagonal one below $300^{\rm o}$C;
(ii) all the samples studied in our work contain   the
superconducting tetragonal phase as the main phase with almost the
same average stoichiometry ($\rm FeSe_{0.974(2)}$) (see insert in
Fig.5 and Table~1) and display the same $\rm T_{c}\simeq 8K$. This
is in disagreement with the results of
Ref.~\onlinecite{McQueenPRB2009} because the compounds with the
stoichiometries shown by points in the inset of
Fig.~\ref{fig:Phase_diagram} would have to display lower ($\sim$5 K)
or even vanishing $T_{\rm c}$;
(iii) the present work demonstrates that there is no need for
quenching from high temperatures (300-450$^{\rm o}$C) in order to
get a stable at room temperature  and pure  tetragonal phase.
According to our NPD studies FeSe$_{0.98}$ (LT098) sample contains,
in total, less than 2\% of impurity phases. Consequently, our data
do not prove an existence of tetragonal - hexagonal phase transition
on cooling at $\rm \sim 300^{\rm o}$C as proposed in
Ref.~\onlinecite{McQueenPRB2009};
(iv)our data suggest very narrow range, or even strictly defined
stoichiometry of the superconducting tetragonal FeSe$\rm_{1-x}$
phase. It looks that the composition of this phase is located
between the fields proposed in both
Ref.~\onlinecite{OkamotoJPhEq1991,McQueenPRB2009}. An additional
confirmation of the correct locus of the tetragonal phase on the
phase diagram comes from investigation of a phase composition of the
LT100 sample (nominally FeSe$_{1.00}$). According to the phase
diagram \cite{OkamotoJPhEq1991} this sample should be in the two
phase region ($\rm \beta FeSe$--$\rm \alpha Fe_{7}Se_{8}$) at room
temperature.  Using a lever rule $\sim$18\% Fe$_7$Se$_8$ would be
expected being in a good agreement  with 16.5\% as found from NPD
data (see Table~\ref{Table:results}).

\subsection{The chemical stability of FeSe}

\begin{figure}[htb]
\begin{center}
\leavevmode
\includegraphics[width=1\linewidth]{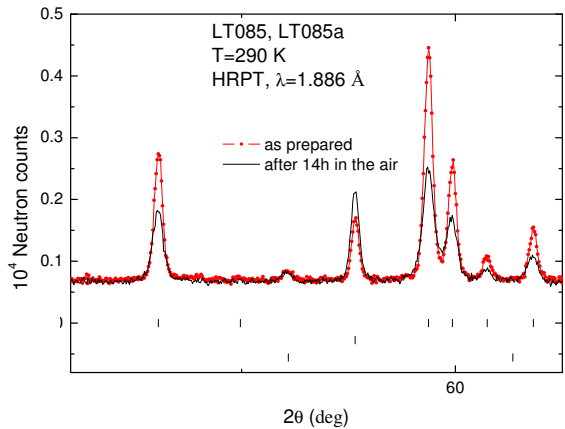}
\end{center}
 \caption{(Color online) Neutron powder diffraction pattern of
LT085 and LT085a samples at T=290~K measured at HRPT with the
wavelength $\lambda=1.886$~{\AA}. The red dotted curve corresponds
to the as prepared sample (LT085). The solid black curve was
obtained after exposing the sample for 14 hours in the air (LT085a).
See text for details.}
 \label{fig:NPD-degraded_sample}
\end{figure}

In order to study  the chemical stability of FeSe-samples  the LT085
sample was powderized and stored in air for 14 hours (LT085a) and
than measured by means of NPD. In Fig.~\ref{fig:NPD-degraded_sample}
the Rietveld refinement of a neutron diffraction pattern of LT085a
sample (solid line) together with as-prepared LT085 sample (dotted
curve) taken at room temperature are presented. The sample underwent
drastic changes after exposing in air. Volume fraction of the main
tetragonal phase was decreased from 93\% down to 84\%, whereas
quantity of Fe increases from 6\% up to 13\%, at the same time the
increase of the of the hexagonal phase content was not so
pronounced. The diffraction peaks of the main phase of LT085a sample
show severe broadening,  and at the same time  the atomic
displacement parameters (Debye-Waller factor)  increase by 1.5
times, thus implying the presence of both large scale defects (like
dislocations or the presence of new-phase particles) and point
defects (e.g. vacancies). \cite{krivoglaz96} Additionally, it was
found that the stoichiometry of the main phase becomes almost 1:1
(Fe:Se). The integral counting rates (scale factors) further reveal,
that about 20-30\% of the main phase was lost  (most probably it
became amorphous).

\begin{figure}[htb]
\begin{center}
\leavevmode
\includegraphics[width=0.9\linewidth]{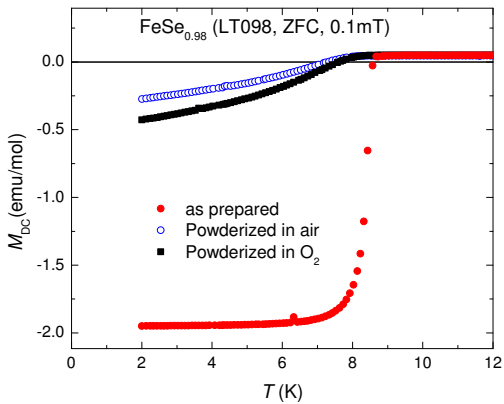}
\end{center}
 \caption{(Color online) Temperature dependencies of the DC
magnetization ($M_{DC}$, zero-field cooling, 0.1~mT) of LT098
sample. The experimental data correspond to the sample: ($\bullet$)
as prepared; ($\circ$) powderized/exposed in the air;
($\blacksquare$) powderized/exposed in pure oxygen. }
 \label{fig:Magn-degraded_sample}
\end{figure}

In order to figure out the reason of FeSe$_{1-x}$ degradation,
additional experiments were performed. The FeSe$_{0.98}$ (LT098)
sample was divided in 3 parts. Each of them was further powderized
and exposed in pure helium, oxygen and air atmosphere.
Figure~\ref{fig:Magn-degraded_sample} shows $M_{DC}(T)$ curves
obtained for the different parts of the sample. It is obvious that
both, air and oxygen, lead to a dramatic degradation of the
superconducting properties. Indeed, the superconducting volume
fraction decreases by more than a factor of 5, while the $T_{\rm c}$
onset shifts to the lower temperature. At the same time the
superconducting transition becomes very broad -- the magnetization
decreases continuously from $T_{\rm c}$ down to 2~K. We suppose,
therefore, that by exposing FeSe sample in the air or in the oxygen
atmosphere it decomposes by oxidizing (most probably forming
SeO$_2$).

\section{Conclusions}

A comparative study of the crystal structure and the magnetic
properties of the  superconductor $\rm FeSe_{1-x}$ synthesized  at
lower temperatures from powders and at higher temperatures from
pieces of metal - was performed. The effect of a starting (nominal)
stoichiometry on a phase purity of the obtained samples and their
superconducting transition temperatures $T_{\rm c}$ was studied.  On
the base of our neutron powder diffraction data we have revised the
Fe-Se concentration phase diagram proposed by
Okamoto\cite{OkamotoJPhEq1991}. In particular, we have found that in
the Fe-Se system a stable phase exhibiting superconductivity at
$T_{\rm c} \sim 8K$ exists in the narrow range of selenium
concentration ($\rm FeSe_{0.974\pm 0.005}$).

As revealed by NPD study, at  $T\sim100$~K  FeSe$\rm_{1-x}$ undergoes
a second order structural phase transition from a tetragonal phase
(space group \emph{P4/nmm}) to an orthorhombic (space group
\emph{Cmma}) on cooling. Fe-Se-Fe bond angles in the $\rm FeSe_4 $
pyramids become different in low temperature phase, whereas the Fe-Se
bond lengths are not changed at the transition.

The chemical stability of FeSe samples exposed in air and pure oxygen
atmosphere was studied. It was found, that after exposing in air the
structure gets many defects, as revealed by  NPD diffraction peaks
broadening  and the large increase in the atomic displacement parameters.
The amount of both impurity phases increases about two times reaching
26\% for metallic iron and 2\% for hexagonal FeSe. This leads to a
dramatic degradation of the superconducting properties, which was
proved by magnetization measurements.

\section*{A{\lowercase{cknowledgements}}}

The authors are grateful to Prof. Hugo Keller for helpful
discussions. This study was partly performed at Swiss neutron
spallation SINQ of Paul Scherrer Institute PSI (Villigen, PSI). We
acknowledge the allocation of the beam time at the HRPT
diffractometer of the Laboratory for Neutron Scattering (ETHZ \&
PSI, Switzerland).  The authors thank the NCCR MaNEP project and the
Swiss National Science Foundation for the support of this study.

\end{document}